\documentclass[sigconf,nonacm]{acmart}
\usepackage{multirow}
\usepackage{subfigure}
 \usepackage{url}
 \usepackage{multirow}

\usepackage{algorithmic}
\usepackage{graphicx}
\usepackage{xcolor}
\usepackage{soul}

\AtBeginDocument{%
  }

\setcopyright{none}

\begin{document}

\title{Soft GPGPU versus IP cores: Quantifying and Reducing the Performance Gap}

\author{Martin Langhammer}
\affiliation{%
\institution{Intel Corporation \& Imperial College London}
\country{UK}
}
\email{martin.langhammer@intel.com}

\author{George A. Constantinides}
\affiliation{%
  \institution{Imperial College London}
\country{UK}
}
\email{g.constantinides@imperial.ac.uk}

\begin{abstract}

eGPU, a recently-reported soft GPGPU for FPGAs, has demonstrated very high clock frequencies (more than 750 MHz) and small footprint. This means that for the first time, commercial soft processors may be competitive for the kind of heavy numerical computations common in FPGA-based digital signal processing. In this paper we take a deep dive into the performance of the eGPU family on FFT computation, in order to quantify the performance gap between state-of-the-art soft processors and commercial IP cores specialized  for this task. In the process, we propose two novel architectural features for the eGPU that improve the efficiency of the design by 50\% when executing the FFTs. The end-result is that our modified GPGPU takes only 3$\times$ the performance-area product of a specialized IP core, yet as a programmable processor is able to execute arbitrary software-defined algorithms. Further comparison to Nvidia A100 GPGPUs demonstrates the superior efficiency of eGPU on FFTs of the size studied (256 to 4096-point).
\end{abstract}

\maketitle

\section{Introduction}

Modern FPGAs have very high arithmetic densities, with many thousands of DSP Blocks, which can support over 15 TFLOPs of IEEE-754 single-precision floating-point (FP32) arithmetic~\cite{Agilex}. Accessing this performance is possible using RTL ({\em{e.g.}} VHDL or Verilog) design methods, but has so far eluded a software-based approach. 

This work builds on our recently published eGPU~\cite{eGPU_FPL}~\cite{ISFPGA_eGPU}, which is a general-purpose SIMT (Single Instruction Multiple Thread) processor optimized for FPGA. The eGPU operates at high clock rate (up to 771 MHz) while requiring modest area (5K-10K LUTs for one streaming mult-processor (SM) comprised of 16 scalar processors (SP)). The cost and performance of the eGPU is at a different level than previously reported soft GPGPUs~\cite{Guppy}~\cite{SCRATCH}~\cite{FlexGrip}~\cite{FlexGrip_Thesis}~\cite{FGPU}~\cite{DOGPU}, which are large (typically 100K to 300K LUTs) and/or feature a low Fmax (30MHz to 300MHz). Although vector processor architectures have been productized for FPGA~\cite{MicrosemiVectorBlox} (based on published works~\cite{Vegas}~\cite{Venice}~\cite{VectorBlox}), the Fmax is low ($\approx$ 150MHz) and floating-point is not supported.

A fixed-function IP will always have the potential for higher performance than a soft CPU. However, for many applications {\em e.g.}~those where multiple algorithmic passes are applied to the same data, especially if those passes are not known in advance of runtime, a high-performance processor is compelling. The soft processor needs no hardware reconfiguration to reuse the underlying FPGA architecture for the different passes, while allowing for easy and efficiently programming using software flows. The processor will not be as performant on any one fixed compute kernel when compared to a customized accelerator, but we have the option of using several processors, especially if they each occupy only $\sim$1\% of the FPGA area.

The FFT is a very common kernel for FPGA-based computation, and this has inspired the work presented in this paper. Our goal in this work is use the FFT benchmark to drive deeper into closing the performance gap between a specialized IP core and the eGPU on a wide range of commercially interesting FFTs design points.

We also propose modifications to the original eGPU design: enhanced memory architectures, and improved FP datapaths that support complex numbers more effectively. These enhancements follow the same approach of the eGPU, {\em i.e.}~an FPGA-first approach: designing the software processor around the FPGA fabric details rather than simply mapping the processor onto an FPGA. We are able to (i) maintain the high clock frequency of the previously published GPGPU core, (ii) while only marginally increasing resource utilisation (taking a balanced approach to the use of logic, DSP, and memory), and (iii) demonstrate that the resulting core packs efficiently into the FPGA, and meets the area and performance goals with a minimum (or none) of design-tool constraints.

We make the following novel contributions:
\begin{enumerate}
\item We implement and profile FFTs from 256-point to 4096-point, for radices 2, 4, 8, and 16, and evaluate them across 6 different GPGPU architectural variants, to build up a comprehensive picture of soft GPGPU architecture and programming.  
\item We describe and implement a GPGPU shared-memory with additional virtual write ports, which offers enhanced performance for applications such as FFTs and reduction. 
\item We describe and implement a GPGPU with complex number functional units. These improve the throughput of algorithms like the FFT, and have a net zero area impact, as they use the FPGA resources more efficiently, with the floorplan essentially unchanged.
\item We introduce a normalized comparison between dedicated FPGA IP, FPGA processor implementations, and the use of dedicated processors such as commercial GPUs.
\end{enumerate}

\section{FFT Solutions}
\label{sec:background}

In addition to implementing and reporting the performance of FFTs on a soft GPGPU, we compare the absolute and relative performance of this programmable soft solution with methods of calculating high performance FFTs. We compare directly to dedicated soft FFT cores~\cite{IntelFFT}, and also compare the efficiency of a soft GPGPU to a commercial GPGPU.

Most of the current FPGA FFT IP cores are streaming, {\em i.e.} after some amount of latency, the transformed input data is output at the same rate as the input, without break~\cite{FFT_Survey}. Throughput performance is easily calculated as the dataset size divided by the clock frequency.

We also compare our soft GPGPU implementation to a recent commercial GPU~\cite{A100}. We will use the metric of efficiency - the percentage of FLOPs realizable. This is a fair comparison, because the FP32 density (per mm\textsuperscript{2}) is similar on contemporary FPGAs and GPGPUs. The recent Nvidia A100-40G provides 19.5 TFLOPs peak with a 826mm$^2$ die size~\cite{A100}. FPGA vendors do not publish their die sizes, however, a mid-range FPGA would be significantly smaller that the 826mm$^2$ of the Nvidia device - after all, there are a number of larger FPGA devices in the Agilex family. For the purposes of this paper, we will assume that the 9.6 TFLOPs on the Agilex AGF022 device is similar in density to the large GPGPU. Although the two devices are on different process nodes, 7nm and 10nm respecitvely, the transistor density is almost the same on the two nodes \cite{wiki_10nm}\cite{wiki_7nm},so the normalized arithmetic density (TFLOPs/mm$^2$) will be roughly the same between the Agilex device and the A100.

Processing a FFT will require a significant amount of data re-ordering between passes, and this will in turn be affected by the memory bandwidth. The GPU can only spend a certain percentage of its cycles on calculating the butterflies, and must spend much of the time organizing data between the threads. Additional considerations include the movement of data across the complex memory hierarchy in the commercial GPUs. 

\subsection{IP vs. Processor Solutions}

Existing work has shown that FPGA IP is able to achieve all of its potential arithmetic performance for a given function~\cite{FFT_Gustafsson}, while the GPU only manages 20\% to 30\%~\cite{cuFFT}. This can be explained by the cost of memory accesses. In each pass of the FFT (this will be explained in more detail in the next section), the entire dataset needs to be reordered, which typically requires transfer through the shared memory. This requires both a read and write of the dataset, during which no processing occurs. The IP core on the other hand, can read from one stage buffer, process the data, and store in the next stage buffer, and do this simultaneously. A processor-based FFT will therefore always be less efficient than the IP core because of the memory operation overheads.

\section{FFTs}
\label{sec:FFTs}

We have designed FFT kernels in four different radices: radix-2, radix-4, radix-8, and radix-16. These kernels are also commonly known as butterflies for radix-2 and dragonflies for radix-4. All of our kernels are themselves implemented with radix-2 butterflies. 

For any given radix FFT, one kernel will be calculated per thread; the results of that kernel are then multiplied by a twiddle factor in the same thread. The FFTs (ranging from 256 to 4096 points), for any given radix, are based on the kernel of the respective radix.

\subsection{Reducing Operations}
\label{reducing_ops}

The FFT is computationally intensive, with 10 flops required per radix-2 butterfly. We can reduce the number of flops somewhat in the higher radix kernels, by observing that some of the twiddle factors (complex coefficients, often expressed as $W$\textsuperscript{x})
are computationally simple rotations, such as $0$, $\pm 1$, $\pm j$.
These can be implemented using integer operators, which also have the added benefit of having a lower power consumption than the floating point equivalents. A multiply by '1' (such as $-j\cdot jx$) can be implemented by a integer addition of $'x+0'$. A FP multiply by '-1' can be implemented using an integer XOR by x"8000". Other twiddle factors have the same coefficient for both the real and imaginary components, so we only need two multiplications, rather than four multiplications and two addition/subtractions.

If we analyse the twiddle factors, a radix-2 16 point FFT (the construction of the DFT (Discrete Fourier Transform) kernel used by our radix-16 FFTs) there are 16 distinct $W$ values, which would normally require 96 flops for the complex multiplies. But if we look at the actual values closely, we can see that we only need four complex multiplies (24 flops), 12 real multiplies, and 14 other arithmetic operations (50 rather than the 96 in the pedantic implementation) which can be implemented using either INT or FP operations, as described above. 

\subsection{Address Generation and Digit-Reversed Indexing}

Generally, the FP operations will be used for calculating the FFT, and the INT operations for configuring the addressing. One result of most FFT algorithms is that the output is not organized in natural order, but rather in digit-reversed order. For example, for radix-2, the natural order can be found by changing the order of the bits of the address during reading the output. For a streaming core, this may require double buffering the output, which adds resources and latency.

As the entire FFT is contained in the GPU registers during processing, we only need to create a new address value in one of the registers to write back the data into shared memory in natural order. We thereby only need a few additional instructions, and no additional (logic, memory, or DSP) resources. The time impact is minimal, as generating a new address is a fraction of the time required to process any given pass.

\subsection{Commercial GPGPU FFT Code}

The geometry of the access patterns can have a significant effect on the memory efficiency of the FFT implementation. Stockham~\cite{StockhamFFT} or Pease~\cite{PeaseFFT} FFTs have constant geometries, which may improve efficiency in avoiding bank conflicts in commercial GPGPU architectures, and are commonly used for these devices~\cite{FFT_Franchetti}. The eGPU uses a somewhat different style of shared memory in order to map to the FPGA architecture efficiently, and we can use the standard Cooley-Tukey access patterns.

\section{Virtual Banked Shared Memory}

\begin{figure*}
    \centering
    \Description{Soft GPGPU with Virtual Multi-Port Memory. Architectural enhancements over the original eGPU are shown on the right hand side.}
    \includegraphics[scale=0.60]{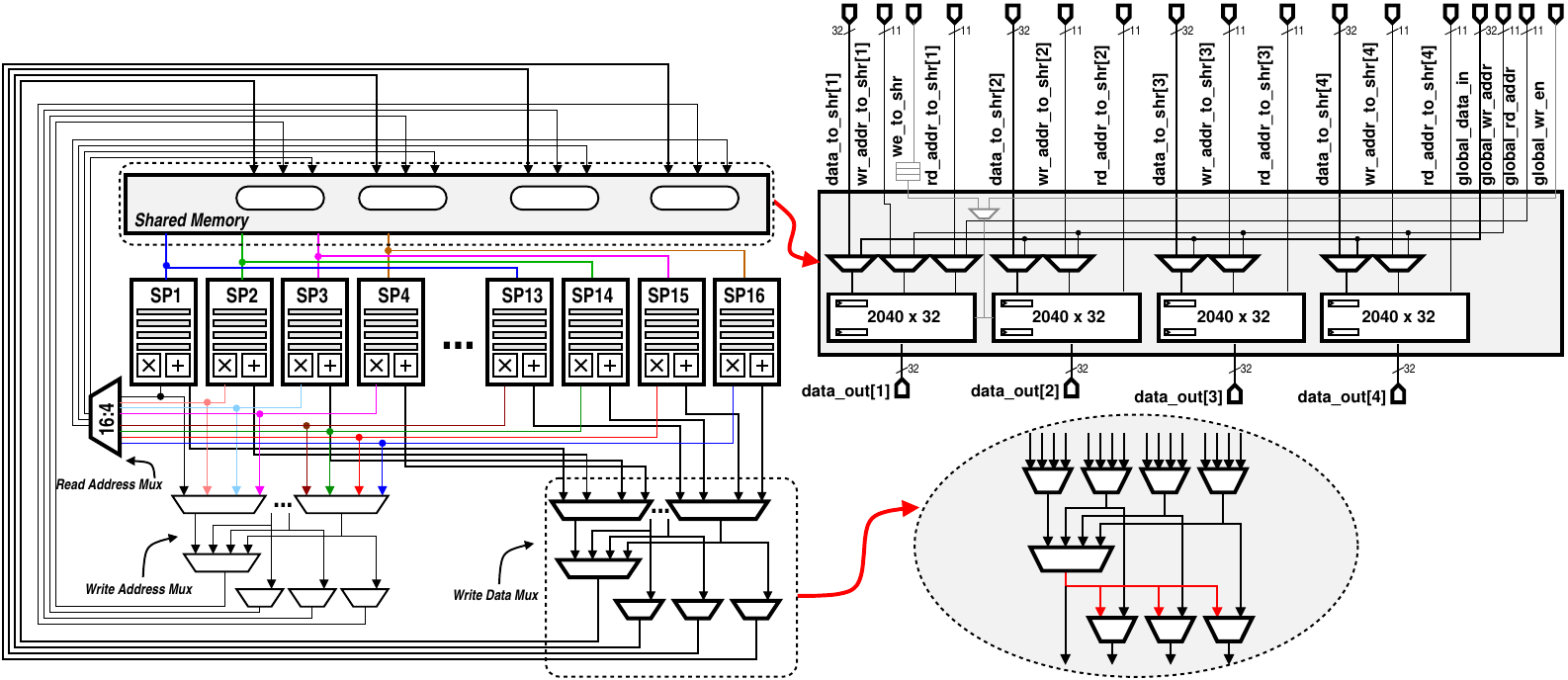}
    \caption{Soft GPGPU with Virtual Multi-Port Memory. Architectural enhancements over the original eGPU are shown on the right hand side.}
    \label{fig:sparse_block}
\end{figure*}

\begin{figure*}
    \centering
    \Description{Mapping}
\begin{minipage}{\textwidth}
\centering
\scriptsize
\begin{tabular}{|@{~}c@{~}|@{~}c@{~}|@{~}c@{~}|@{~}c@{~}|@{~}c@{~}|@{~}c@{~}|@{~}c@{~}|@{~}c@{~}|@{~}c@{~}|@{~}c@{~}|@{~}c@{~}|@{~}c@{~}|@{~}c@{~}|@{~}c@{~}|@{~}c@{~}|@{~}c@{~}|@{~}c@{~}|@{~}c@{~}|@{~}c@{~}|@{~}c@{~}|@{~}c@{~}|@{~}c@{~}|@{~}c@{~}|@{~}c@{~}|@{~}c@{~}|@{~}c@{~}|@{~}c@{~}|@{~}c@{~}|@{~}c@{~}|@{~}c@{~}|@{~}c@{~}|@{~}c@{~}|@{~}c@{~}|}
\hline
\multirow{5}{*}{\rotatebox{90}{Pass 1}} &  {T0} & {T1} & {T2} & {T3} & {T4} & {T5} & {T6} & {T7} & {T8} & {T9} & {T10} & {T11} & {T12} & {T13} & {T14} & {T15} & {T16} & {T17} & {T18} & {T19} & {T20} & {T21} & {T22} & {T23} & {T24} & {T25} & {T26} & {T27} & {T28} & {T29} & {T30} & {T31} \\
      \cline{2-33}
 &  {i000} & {i001} & {i002} & {i003} & {i004} & {i005} & {i006} & {i007} & {i008} & {i009} & {i010} & {i011} & {i012} & {i013} & {i014} & {i015} & {i016} & {i017} & {i018} & {i019} & {i020} & {i021} & {i022} & {i023} & {i024} & {i025} & {i026} & {i027} & {i028} & {i029} & {i030} & {i031}\\
       \cline{2-33}
  &  {i064} & {i065} & {i066} & {i067} & {i068} & {i069} & {i070} & {i071} & {i072} & {i073} & {i074} & {i075} & {i076} & {i077} & {i078} & {i079} & {i080} & {i081} & {i082} & {i083} & {i084} & {i085} & {i086} & {i087} & {i088} & {i089} & {i090} & {i091} & {i092} & {i093} & {i094} & {i095} \\
        \cline{2-33}
   &  {i128} & {i129} & {i130} & {i131} & {i132} & {i133} & {i134} & {i135} & {i136} & {i137} & {i138} & {i139} & {i140} & {i141} & {i142} & {i143} & {i144} & {i145} & {i146} & {i147} & {i148} & {i149} & {i150} & {i151} & {i152} & {i153} & {i154} & {i155} & {i156} & {i157} & {i158} & {i159} \\
         \cline{2-33}
    &  {i192} & {i193} & {i194} & {i195} & {i196} & {i197} & {i198} & {i199} & {i200} & {i201} & {i202} & {i203} & {i204} & {i205} & {i206} & {i207} & {i208} & {i209} & {i210} & {i211} & {i212} & {i213} & {i214} & {i215} & {i216} & {i217} & {i218} & {i219} & {i220} & {i221} & {i222} & {i223} \\
    \hline
\end{tabular}
\vskip 0.2cm
\begin{tabular}{|@{~}c@{~}|@{~}c@{~}|@{~}c@{~}|@{~}c@{~}|@{~}c@{~}|@{~}c@{~}|@{~}c@{~}|@{~}c@{~}|@{~}c@{~}|@{~}c@{~}|@{~}c@{~}|@{~}c@{~}|@{~}c@{~}|@{~}c@{~}|@{~}c@{~}|@{~}c@{~}|@{~}c@{~}|@{~}c@{~}|@{~}c@{~}|@{~}c@{~}|@{~}c@{~}|@{~}c@{~}|@{~}c@{~}|@{~}c@{~}|@{~}c@{~}|@{~}c@{~}|@{~}c@{~}|@{~}c@{~}|@{~}c@{~}|@{~}c@{~}|@{~}c@{~}|@{~}c@{~}|@{~}c@{~}|}
\hline
\multirow{5}{*}{\rotatebox{90}{Pass 2}} &  {T0} & {T1} & {T2} & {T3} & {T4} & {T5} & {T6} & {T7} & {T8} & {T9} & {T10} & {T11} & {T12} & {T13} & {T14} & {T15} &  {T16} & {T17} & {T18} & {T19} & {T20} & {T21} & {T22} & {T23} & {T24} & {T25} & {T26} & {T27} & {T28} & {T29} & {T30} & {T31} \\
     \cline{2-33}
  & {i000} & {i001} & {i002} & {i003} & {i004} & {i005} & {i006} & {i007} & {i008} & {i009} & {i010} & {i011} & {i012} & {i013} & {i014} & {i015} & {i064} & {i065} & {i066} & {i067} & {i068} & {i069} & {i070} & {i071} & {i072} & {i073} & {i074} & {i075} & {i076} & {i077} & {i078} & {i079} \\
       \cline{2-33}
  & {i016} & {i017} & {i018} & {i019} & {i020} & {i021} & {i022} & {i023} & {i024} & {i025} & {i026} & {i027} & {i028} & {i029} & {i030} & {i031} & {i080} & {i081} & {i082} & {i083} & {i084} & {i085} & {i086} & {i087} & {i088} & {i089} & {i090} & {i091} & {i092} & {i093} & {i094} & {i095} \\
        \cline{2-33}
   & {i032} & {i033} & {i034} & {i035} & {i036} & {i037} & {i038} & {i039} & {i040} & {i041} & {i042} & {i043} & {i044} & {i045} & {i046} & {i047} & {i096} & {i097} & {i098} & {i099} & {i100} & {i101} & {i102} & {i103} & {i104} & {i105} & {i106} & {i107} & {i108} & {i109} & {i110} & {i111} \\
         \cline{2-33}
    & {i048} & {i049} & {i050} & {i051} & {i052} & {i053} & {i054} & {i055} & {i056} & {i057} & {i058} & {i059} & {i060} & {i061} & {i062} & {i063} & {i112} & {i113} & {i114} & {i115} & {i116} & {i117} & {i118} & {i119} & {i120} & {i121} & {i122} & {i123} & {i124} & {i125} & {i126} & {i127}\\
    \hline
    \end{tabular}
\vskip 0.2cm
\begin{tabular}{|@{~}c@{~}|@{~}c@{~}|@{~}c@{~}|@{~}c@{~}|@{~}c@{~}|@{~}c@{~}|@{~}c@{~}|@{~}c@{~}|@{~}c@{~}|@{~}c@{~}|@{~}c@{~}|@{~}c@{~}|@{~}c@{~}|@{~}c@{~}|@{~}c@{~}|@{~}c@{~}|@{~}c@{~}|@{~}c@{~}|@{~}c@{~}|@{~}c@{~}|@{~}c@{~}|@{~}c@{~}|@{~}c@{~}|@{~}c@{~}|@{~}c@{~}|@{~}c@{~}|@{~}c@{~}|@{~}c@{~}|@{~}c@{~}|@{~}c@{~}|@{~}c@{~}|@{~}c@{~}|@{~}c@{~}|}
\hline
\multirow{5}{*}{\rotatebox{90}{Pass 3}} &  {T0} & {T1} & {T2} & {T3} & {T4} & {T5} & {T6} & {T7} & {T8} & {T9} & {T10} & {T11} & {T12} & {T13} & {T14} & {T15} &  {T16} & {T17} & {T18} & {T19} & {T20} & {T21} & {T22} & {T23} & {T24} & {T25} & {T26} & {T27} & {T28} & {T29} & {T30} & {T31} \\
     \cline{2-33}
  & {i000} & {i001} & {i002} & {i003} & {i016} & {i017} & {i018} & {i019} & {i032} & {i033} & {i034} & {i035} & {i048} & {i049} & {i050} & {i051} & {i064} & {i065}  & {i066}  & {i067}  & {i080}  & {i081}  & {i082}  & {i083} & {i096} & {i097} & {i098} & {i099} & {i112} & {i113} & {i114} & {i115}\\
       \cline{2-33}
  & {i004} & {i005} & {i006} & {i007} & {i020} & {i021} & {i022} & {i023} & {i036} & {i037} & {i038} & {i039} & {i052} & {i053} & {i054} & {i055} & {i068} & {i069}  & {i070}  & {i071}  & {i084}  & {i085}  & {i086}  & {i087} & {i100} & {i101} & {i102} & {i103} & {i116} & {i117} & {i118} & {i119}\\
        \cline{2-33}
  & {i008} & {i009} & {i010} & {i011} & {i024} & {i025} & {i026} & {i027} & {i040} & {i041} & {i042} & {i043} & {i056} & {i057} & {i058} & {i059} & {i072} & {i073}  & {i074}  & {i075}  & {i088}  & {i089}  & {i090}  & {i091} & {i104} & {i105} & {i106} & {i107} & {i120} & {i121} & {i122} & {i123}\\
         \cline{2-33}
  & {i012} & {i013} & {i014} & {i015} & {i028} & {i029} & {i030} & {i031} & {i044} & {i045} & {i046} & {i047} & {i060} & {i061} & {i062} & {i063} & {i076} & {i077}  & {i078}  & {i079}  & {i092}  & {i093}  & {i094}  & {i095} & {i108} & {i109} & {i110} & {i111} & {i124} & {i125} & {i126} & {i127}\\ 
    \hline
    \end{tabular}

\end{minipage}
    
    \caption{Data Indexes per Pass. The first 31 threads of Passes 1, 2, and 3 are shown, with the thread number in the first register, followed by data indexes of the complex component in the next four registers.}
    \label{fig:passes}
\end{figure*}

A representative architecture of the eGPU, depicting 16 streaming processors (SP) is shown in Figure~\ref{fig:sparse_block}. This single streaming multiprocessor (SM) can support a wide range of parameters, including thread size (up to 4096 threads), registers per thread (1024), and shared memory size (256KB). Memory bandwidth was one of the major performance limiting factors on the eGPU~\cite{eGPU_FPL}. 

One of the methods that commercial GPUs use to increase local memory bandwidth is by organizing the memory in banks~\cite{shared_banks}. The large and complex logic required for bank arbitration, plus the wire load needed for the extensive bus routing, can make this option very expensive in the context of the compact eGPU architecture.  

Instead, we introduce a manually (software) controlled bank system, where we can optionally write four independent values into the four parallel memory banks. The original eGPU uses a 4R-1W (4 read ports and 1 write port) shared memory - all 4 banks contain the same data. Our virtual banked memory has a new instruction (\textbf{save\_bank}) which writes values from 4 SPs simultaneously. This means that for 
any memory location written to by the banked instruction, three of the memory banks will contain invalid information. We can therefore only use this feature when we know the next read to that location will map that shared memory bank to the SP index modulo 4, {\em{i.e.}} memory bank 1 maps to SP1, 5, 9 and 13, memory bank 2 maps to SP 2, 6, 10 and 14, and so on. It turns out that many algorithms can use this approach (the description and analysis of these are beyond the scope of this paper). We can also see this in a regular GPGPU  - there are many cases where little arbitration is required – in the case of the eGPU, we need to manually invoke the writing. The read operation – both in use and in architecture - is unchanged.

The manually banked eGPU can support the current (where all four banks contain the same information at any given location), and the new (where only one of the four banks contain valid information at any given location) storage at the same time. Some memory ranges will contain one format, and other ranges (such as the FFT working data) will contain the new format. The FFT working data locations may switch from the new to the standard format from pass to pass – the only difference is in the longer write times (4$x$ the number of cycles).

We will illustrate the workings of the virtual banks using Figure~\ref{fig:passes} for a radix-4, 256 point FFT. Each column shows the first 5 registers of each thread (R0 contains the thread number, such as T7 for thread 7), and the next 4 rows for each thread show the index of the FFT pass data. The 32 depicted columns correspond to the first 32 threads (out of 64) initialized for this FFT. Comparing each column of Pass 1 with Pass 2, we can see that the same indexes are found in the same SPs -- Pass 2 T0 requires indexes 0, 16, 32 and 48, which are found in the first SP for both passes. Comparing Pass 2 with Pass 3, we see that the required indexes for Pass 3 are found in the SPs modulo 4 in Pass 2. For example Pass 3 T0 requires indexes 0, 4, 8 and 12 which are found in SPs 1, 5, 9 and 13 in Pass 2. A write to virtual banks in Pass 2 will place all these in the first memory bank. But we cannot use these scheme between Pass 3 and Pass 4. Pass 4 T0 will require indexes 0, 1, 2, 3 which will be written to different memory banks using the virtual bank write from Pass 3. We therefore need to revert to the original memory format. The only change to the FFT program is the write instruction (\textbf{save} instead of \textbf{save\_bank}), and only has a time impact (the write cycle is 4x as long).

\section{Complex Operators}
\label{sec:complex}

The FFT, like many other types of signal processing algorithms, uses many complex multiplies. In most processors, these are performed by multiple real operations (a complex multiply consists of four real multiplies, an addition and a subtraction). The FPGA is rich in both embedded DSP Blocks and embedded memories, but limited in connectivity. In particular, memory ports are limited. We could build a four read port register file from M20Ks (by using four memories in parallel in DP mode, or two in parallel in the slower QP mode), but this would be expensive in terms of memory, increase power consumption, and the additional 32-bit busses from the memory columns to the DSP columns would potentially impact the fitting of the eGPU into the FPGA.

As the twiddle factors are the same for both the real and imaginary calculations, we could load the twiddle factor into the FP ALU in the first cycle, and then calculate the two outputs on the following cycle. This, however, is only possible if the wavefront (the number of cycles that each instruction is run for in the current thread initialization) is one cycle deep - in a typical case, we would be calculating multiple results per instruction, with a depth of multiple cycles.

\begin{figure}
    \centering
    \Description{Complex}
    \rotatebox{-90}{\includegraphics[scale=0.70]{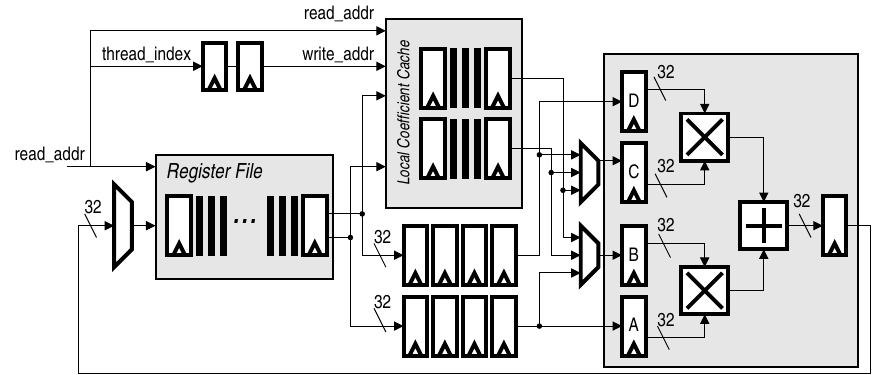}}
    \caption{Complex Functional Unit, Including New Coefficient Cache.}
    \label{fig:complex}
\end{figure}

Instead, we introduce a local coefficient cache, which is shown in Figure~\ref{fig:complex}. A new instruction (\textbf{lod\_coeff}) loads the contents of two registers (containing the twiddle factors) into a cache implemented as a circular buffer. Both the write and read addresses for the cache are the thread index (in the case of the write address, this is delayed by two cycles to align the output of the register file). As the cache only holds a single (complex) value per thread at a time, it will be a fraction of the size of the register file. The figure shows a number of registers between the register file and the FP functional unit. These represent the total pipeline delay of the SP, and may be distributed in other locations between the output and the input of the register file. 

An example complex multiplication is invoked as follows. One of the complex operands (for example the the output of the lower arm of a radix-2 butterfly) is in R8 (real component) and R9 (imaginary component), and the other operand (for example the twiddle factor) is in R30 and R31.

\begin{verbatim}
LOD_COEFF R30, R31; -- load tw_real, tw_imag into cache
MUL_REAL R6, R8, R9; -- R6 = (R8 * tw_real) - (R9 * tw_imag) 
MUL_IMAG R7, R8, R9; -- R7 = (R8 * tw_imag) + (R9 * tw_real)
\end{verbatim}

The sum of two multipliers FP functional unit is constructed from two DSP Blocks in the Agilex device. This structure also supports the standard floating-point operations (multiply, add and subtract). The standard FP multiply routes the two register busses on functional unit ports A and B, and '0' on C and D. A FP addition routes the two register busses to ports A and D, and a '1' to ports B and C. A subtraction is the same as the addition, other than the MSB (FP sign bit) is inverted on port D. Note that not all mux combinations are shown in Figure~\ref{fig:complex} for clarity. The cache will run off the thread index continuously, whether used or not, which will increase power consumption slightly. We have added a pair of instructions to enable and disable the cache (\textbf{coeff\_en}, \textbf{coeff\_dis}), which gate the clock enable.

As the cache is significantly smaller than the register files, it will most likely be mapped to MLABs (using an unconstrained compile, however, the memory target type can also be specified in the memory instantiation by a parameter if needed), and have a minimal impact on the fitting of the eGPU. Although the number of 32-bit busses into the functional unit doubles over the standard unit (single operation), an MLAB-based cache has much more flexibility in fitting as there are multiple logic columns around the DSP columns, making it much easier to find a connection.

The number of DSP Blocks is doubled per eGPU to support the complex mode (the integer multiplier is not required for FFTs), but the overall cost (measured in die area consumed) remains the same. This is because the ratio of DSP Blocks to ALMs is still somewhat less than the ratio provided per local area (sector) in the FPGA. (If an unused DSP Block is surrounded by logic, it will not be otherwise available to other circuits on that FPGA.)

\section{Results}

\begin{table*}
\footnotesize
  \begin{center}
    \caption{Radix-4 FFT Profiling - Cycles per Operation and Performance}
    \label{tab:r4_fft}
    \begin{tabular}{|c|c||c|c|c|c|c|c||}
\hline
  \textbf{Points} & \textbf{Type} & \textbf{eGPU-DP} & \textbf{eGPU-DP} & \textbf{eGPU-DP}  & \textbf{eGPU-DP} & \textbf{eGPU-QP} & \textbf{eGPU-QP} \\
    \ & & &  \textbf{VM}  & \textbf{Complex} & \textbf{VM+Complex} & & \textbf{Complex} \\ 
      \hline\hline
      \multirow{5}{*}{4096} & {FP OP} & {13440} & {13440} & {7680} & {7680} & {13440} & {7680}   \\
    \ & {Complex OP} & {-} & {-} & {2880} & {2880} & {-} & {2880}   \\
    \ & {INT OP} & {2880} & {2880} & {2880} & {2880} & {2880} & {2880}   \\
    \ & {Load} & {19968} & {19968} & {19968} & {19968} & {19968} & {19968} \\
    \ & {Store} & {49152} & {16384} & {49152} & {16384} & {24576} & {24576} \\
    \ & {StoreVM} & {-} & {8192} & {-} & {8192} & {-} & {-}\\
    \ & {Immediate} & {1287} & {1287} & {1287} & {1287} & {1287} & {1287} \\
    \ & {Branch} & {90} & {90} & {90} & {90} & {90} & {90} \\   
    \cline {2 -8}
    \ & {Total} & {86817} & {62214} & {83937} & {59361} & {62241}  & {59361} \\
    \ & {Time (us)} & {112.60} & {80.73} & {108.87} & \bf{76.99} & {103.74}  & {98.94} \\
    \ & {Efficiency~\%} & {15.48} & {21.60} & {15.82} & \bf{22.64} & {21.59} & \bf{22.64} \\
    \ & {Memory~\%} & {79.61} & {71.59} & {82.35} & {75.04} & {71.56} & {75.03} \\
    \hline
      \multirow{5}{*}{1024} & {FP OP} & {2752} & {2752} & {1600} & {1600} & {2752} & {1600}   \\
    \ & {Complex OP} & {-} & {-} & {576} & {576} & {-} & {576}   \\
    \ & {INT OP} & {576} & {576} & {576} & {576} & {576} & {576}  \\
    \ & {Load} & {4096}  & {4096}  & {4096}  & {4096}  & {4096}  & {4096} \\
    \ & {Store} & {10240} & {4096} & {10240} & {4096} & {5120} & {5120}\\
    \ & {StoreVM} & {-} & {1536} & {-} & {1536} & {-} & {-}\\
    \ & {Immediate} & {262} & {262} & {262} & {262} & {262} & {262}\\
    \ & {Branch} & {114} & {114} & {114} & {114} & {114}  & {114}  \\     \cline {2 -8}
    \ & {Total} & {18040} & {13432} & {17464} & {12856} & {12920} & {12344} \\
    \ & {Time (us)} & {23.40} & {17.42} & {22.65} & \bf{16.67} & {21.53}  & {20.57} \\
    \ & {Efficiency} & {15.25} & {20.49} & {15.76} & {21.41} & {21.30} & \bf{22.29} \\
        \ & {Memory~\%} & {79.47} & {72.42} & {82.09} & {75.67} & {71.33} & {74.66} \\
    \hline
    \multirow{5}{*}{256} & {FP OP} & {536} & {536} & {320} & {320} & {536} & {320}   \\ 
     \ & {Complex OP} & {-} & {-} & {108} & {108} & {-} & {108}   \\
    \ & {INT OP} & {108} & {108} & {108} & {108} & {108} & {108}  \\
    \ & {Load} & {800}  & {800}  & {800}  & {800}  & {800}  & {800} \\
    \ & {Store} & {2048} & {1024} & {2048} & {1024} & {1024} & {1024}\\
        \ & {StoreVM} & {-} & {256} & {-} & {256} & {-} & {-}\\
    \ & {Immediate} & {76} & {76} & {67} & {67} & {} & {67}\\
    \ & {Branch} & {78} & {78} & {78} & {78} & {78} & {78} \\
    \ & {NOP} & {493} & {493} & {79} & {79} & {301} & {79} \\    
    \cline {2 -8}
    \ & {Total} & {4193} & {3371} & {3608} & {2840} & {2847} & {2584} \\
    \ & {Time (us)} & {5.44} & {4.37} & {4.68} & \bf{3.68} & {4.75} & {4.31} \\
    \ & {Efficiency} & {12.78} & {15.90} & {14.86} & {18.87} & {18.48} & \bf{20.74} \\
     \ & {Memory~\%} & {67.92} & {61.70} & {78.94} & {73.24} & {64.07} & {70.59} \\
    \hline
    \end{tabular}
  \end{center}
\end{table*}

\begin{table*}
\footnotesize
  \begin{center}
    \caption{Radix-8 FFT Profiling - Cycles per Operation and Performance}
    \label{tab:r8_fft}
    \begin{tabular}{|c|c||c|c|c|c|c|c||}
\hline
  \textbf{Points} & \textbf{Type} & \textbf{eGPU-DP} & \textbf{eGPU-DP} & \textbf{eGPU-DP}  & \textbf{eGPU-DP} & \textbf{eGPU-QP} & \textbf{eGPU-QP} \\
    \ & & &  \textbf{VM}  & \textbf{Complex} & \textbf{VM+Complex} & & \textbf{Complex} \\ 
      \hline\hline
      \multirow{5}{*}{4096} & {FP OP} & {11840} & {11840} & {7808} & {7808} & {11840} & {7808}   \\ 
     \ & {Complex OP} & {-} & {-} & {2016} & {2016} & {-} & {2016}   \\
    \ & {INT OP}  & {3296} & {3296} & {2720} & {2720} & {3296} & {2720}  \\
    \ & {Load}   & {13568} & {13568} & {13568} & {13568} & {13568} & {13568} \\
    \ & {Store}   & {32768} & {16384} & {32768} & {16384} & {16384} & {16384} \\
    \ & {StoreVM} & {-} & {4096} & {-} & {4096} & {-} & {-}\\
    \ & {Immediate}  & {328} & {328} & {343} & {343} & {328} & {343} \\  
    \cline {2 -8}
    \ & {Total} &  {61896} & {49608} & {59319} & {47031} & {45512} & {42935} \\
    \ & {Time (us)} & {80.28} & {64.34} & {76.94} & \bf{61.00} & {75.85}  & {71.56} \\
    \ & {Efficiency} & {19.13} & {23.87} & {19.96} & {25.17} & {26.02} & \bf{27.57}\\
    \ & {Memory(\%)} & {74.86} & {68.63} & {78.11} & {72.39} & {65.81} & {69.76} \\
    \hline
      \multirow{5}{*}{512} & {FP OP} & {1068} & {1068} & {732} & {732} & {1068} & {732}   \\ 
     \ & {Complex OP} & {-} & {-} & {168} & {168} & {-} & {168}   \\
    \ & {INT OP}  & {284} & {284} & {236} & {236} & {284} & {236}  \\
    \ & {Load}   & {1216} & {1216} & {1216} & {1216} & {1216} & {1216} \\
    \ & {Store}   & {3072} & {2048} & {3072} & {2048} & {1536} & {1536} \\
    \ & {StoreVM} & {-} & {256} & {-} & {256} & {-} & {-}\\
    \ & {Immediate}  & {40} & {40} & {40} & {40} & {40} & {40} \\
    \ & {NOP}  & {81} & {81} & {81} & {81} & {40} & {40} \\    
    \cline {2 -8}
    \ & {Total} &  {5827} & {5059} & {5779} & {5011} & {4250} & {4202} \\
    \ & {Time (us)} & {7.56} & {6.56} & {7.50} & \bf{6.50} & {7.08}  & {7.00} \\
   \ & {Efficiency} & {18.32} & {21.11} & {18.48} & {21.31} & {25.13} & \bf{25.42} \\
   \ & {Memory(\%)} & {73.59} & {69.58} & {74.20} & {70.25} & {64.75} & {65.49} \\
    \hline
    \end{tabular}
  \end{center}
\end{table*}

\begin{table*}
\footnotesize
  \begin{center}
    \caption{Radix-16 FFT Profiling - Cycles per Operation and Performance}
    \label{tab:r16_fft}
    \begin{tabular}{|c|c||c|c|c|c|c|c||}
\hline
  \textbf{Points} & \textbf{Type} & \textbf{eGPU-DP} & \textbf{eGPU-DP} & \textbf{eGPU-DP}  & \textbf{eGPU-DP} & \textbf{eGPU-QP} & \textbf{eGPU-QP} \\
  \ & & &  \textbf{VM}  & \textbf{Complex} & \textbf{VM+Complex} & & \textbf{Complex} \\ 
      \hline\hline
      \multirow{5}{*}{4096} & {FP OP} & {12384} & {12384} & {6912} & {6192} & {12384} & {6192}   \\ 
    \ & {Complex OP} & {-} & {-} & {2880} & {2880} & {-} & {2880}   \\
    \ & {INT OP} & {1968} & {1968} & {1968} & {1968} & {1968} & {1968}   \\
    \ & {Load} & {9984} & {9984} & {9984} & {9984} & {9984} & {9984} \\
    \ & {Store} & {24576} & {12288} & {24576} & {12288} & {16384} & {16384} \\
    \ & {StoreVM} & {-} & {2048} & {-} & {2048} & {-} & {-}\\
    \ & {Immediate} & {196} & {196} & {154} & {64} & {154} & {64} \\ 
    \cline {2 -8}
    \ & {Total} & {49186} & {38946} & {46552} & {35502} & {40952}  & {37550}  \\
    \ & {Time (us)} & {63.80} & {50.51} & {60.38} & \bf{46.05} & {68.25}  & {62.58} \\
    \ & {Efficiency(\%)} & {25.18} & {31.80} & {27.22} & \bf{35.69} & {30.24} & {33.75} \\
    \ & {Memory(\%)} & {70.26} & {62.45} & {74.24} & {68.50} & {64.39} & {70.22} \\
    \hline
      \multirow{5}{*}{1024} & {FP OP} & {2624} & {2624} & {1472} & {1472} & {2624} & {1472}   \\ 
    \ & {Complex OP} & {-} & {-} & {600} & {600} & {-} & {600}   \\
    \ & {INT OP} & {392} & {392} & {392} & {392} & {392} & {392}  \\
    \ & {Load} & {2496}  & {2496}  & {2496}  & {2496}  & {2496}  & {2496} \\
    \ & {Store} & {6144} & {4096} & {6144} & {4096} & {3072} & {3072}\\
    \ & {StoreVM} & {-} & {512} & {-} & {512} & {-} & {-}\\
    \ & {Immediate} & {143} & {147} & {25} & {25} & {143} & {25}\\
    \cline {2 -8}
    \ & {Total} & {11961} & {10413} & {11290} & {9755} & {8889} & {8219} \\
    \ & {Time (us)} & {15.51} & {13.51} & {14.64} & \bf{12.65} & {14.82}  & {13.70} \\
    \ & {Efficiency} & {21.94} & {25.20} & {23.67} & {27.40} & {29.52} & \bf{32.51} \\
   \ & {Memory(\%)} & {72.23} & {68.07} & {76.53} & {72.82} & {62.64} & {67.75} \\
    \hline
      \multirow{5}{*}{256} & {FP OP} & {486} & {-} & {288} & {-} & {486} & {288}   \\ 
    \ & {Complex OP} & {-} & {-} & {105} & {-} & {-} & {105}   \\
    \ & {INT OP} & {72} & {-} & {72} & {-} & {72} & {72}  \\
    \ & {Load} & {376}  & {-}  & {376}  & {-}  & {376}  & {376} \\
    \ & {Store} & {1024} & {-} & {1024} & {-} & {512} & {512}\\
    \ & {Immediate} & {74} & {-} & {16} & {-} & {74} & {16}\\
    \ & {NOP} & {132} & {-} & {29} & {-} & {132} & {29} \\    
    \cline {2 -8}
    \ & {Total} & {2216} & {-} & {1962} & {-} & {1704} & {1450} \\
    \ & {Time (us)} & {2.87} & {-} & {2.54} & {-} & {2.84}  & \bf{2.42} \\
    \ & {Efficiency} & {21.93} & {-} & {25.38} & {-} & {28.51} & \bf{34.34} \\
    \ & {Memory(\%)} & {63.18} & {-} & {71.36} & {-} & {52.11} & {61.24} \\
    \hline
    \end{tabular}
  \end{center}
\end{table*}

We report three different dataset sizes - 256, 1024, and 4096 points, across 6 different variants of eGPU. One variant includes a 4R-2W shared memory, which uses the M20K blocks in quad-port (QP) mode. This doubles the memory write bandwidth (and reduces the number of M20K blocks needed to construct a given memory size). 

\begin{enumerate}
    \item eGPU-DP: the standard architecture~\cite{eGPU_FPL}, with a 4R-1W shared memory architecture. High performance, at 771MHz.  
    \item eGPU-QP: the standard architecture with a 4R-2W shared memory. The memory style used reduces the performance to 600MHz.
    \item eGPU-DP-VM: the standard eGPU, but with a virtually banked 4R-4W shared memory. 
    \item eGPU-DP-Complex: the standard eGPU with complex multiplier capability.
    \item eGPU-DP-VM-Complex: eGPU with both virtually banked 4R-4W memory and complex multiplier support.
    \item eGPU-QP-Complex: eGPU with 4R-2W shared memory and complex multiplier support.
\end{enumerate}

All eGPU variants used here have the same shared memory size of 64 KB, and a total of 32K registers across all the SPs. We use the following configuration for the FFT tests: 1024 threads with 32 register per thread for the radix-4 tests, and 512 threads with 64 registers per thread for the radix-8 and radix-16 experiments. The selected number of register per thread ensures that all radix-related computations can be preformed without spillage. 
The eGPU-DP variant required 8801 ALMs, 192 M20Ks, and 32 DSP Blocks. The QP version used about half the number of M20Ks. There was a negligible soft logic and memory resource impact in supporting either the the VM or complex features. Although the complex feature added one DSP Block to each SP, we will see that this did not affect the footprint (the total placed and routed area) of the eGPU on the FPGA. The virtual 4R-4W memory is not supported for the eGPU-QP variants, as all memory ports are available for all memory accesses. 

We profiled and reported all variants for both radix-4 (Table~\ref{tab:r4_fft}) and radix-16 (Table~\ref{tab:r16_fft}). With the exception of the 1024 point case for radix-16, these all operate with a single radix, which allows for a straightforward comparison. We also report radix-8 (Table~\ref{tab:r8_fft}) in for 512 and 4096 points. The fastest (time) and most efficient result for each radix/length combination is highlight in boldface. The radix-2 FFTs have lower performance and efficiency than the other radices (largely because of the large number of dataset reordering transfers through shared memory), so we do not report them for brevity. We wrote all of the FFT programs in assembler. The eGPU instruction set~\cite{ISFPGA_eGPU} is similar to the Nvidia PTX ISA~\cite{Nvidia_PTX}.

We benchmark efficiency - the percentage of time that the processor is calculating the FFT ($i.e.$ FP operations) - to compare the different FFT programs, both across multiple eGPU variants as well as other solutions. The peak efficiency is up to around 35\% when both memory bandwidth enhancements (either the 4R-2W or the virtual 4R-4W memory is used), and the complex multiplier support is enabled. We run all of the benchmark FFTs in single-batch mode.

We can see that memory accesses (to and from the shared memory) make up the majority of the cycles. Increasing the memory bandwidth would therefore have the largest impact on performance, but memory bandwidth (the number of read and write ports) is very expensive. In an FPGA, we are limited by the general purpose embedded memory blocks. We can increase the read bandwidth by duplicating memory blocks, in which case the memory cost is directly proportional to the number of ports. We have very little flexibility in increasing the write ports with this style of memory architecture. We can use the embedded M20K blocks in QP mode (where two read and two write ports are emulated inside the memory block), but this reduces the performance of the eGPU to 600MHz. The VM architecture can increase the number of write ports to four, but can only be used for a subset of the FFT passes. 

The twiddle loads accounts for about 10\% of all memory accesses, which we can calculate as follows: In the DP variant, we know that the number of reads will be 25\% of the number of writes based on the 4R-1W architecture. For the 4096 points, radix-16, eGPU-DP example, we have 9984 loads, and 24576 stores. We will need 24576/4 = 6144 data loads, leaving 3840 twiddle loads, or (3840/(24576+9984)) = 11.1\%. We need to bear this cost for the single batch workloads we are running, but these would be amortized away for multi-batch FFTs, increasing the performance by 8\% for the base case. The impact for the higher write bandwidth cases (VM and QP shared memories) would be even higher.

The short pipeline depth (8 cycles) of the eGPU means that hazards are hidden completely if the wavefront depth is greater than 8. With 16 SPs, the wavefront depth can be calculated as $\text{points}/(16 \times \text{radix})$. We only need to insert NOPs to avoid hazards for the shorter FFT examples for all three reported radices.

The use of the complex multiplier feature reduces the number of cycles required for FP operations by about 25\%. Three complex operations are required to complete a complex multiplication: the first one loads the coefficient buffer, and one operation each is required for the real and imaginary results. As the memory accesses still require the majority of the cycles in the FFT, this 25\% FP operation reduction translates into a $\approx$ 5\% performance increase. 

The computational efficiency of the presented FFTs is actually slightly higher than indicated by the number of FP operations, as described earlier in Section~\ref{reducing_ops}. We now profile the radix-8 butterfly (Table~\ref{tab:r8_butterfly}) to show that some of the FFT computation is performed by integer operations. 

\begin{table}
\footnotesize
  \begin{center}
    \caption{Radix-8 Butterfly}
    \label{tab:r8_butterfly}
    \begin{tabular}{|c||c|c|c|c|c||}
\hline
  \textbf{Pass} & \textbf{Operation} & \textbf{Cycle} & \textbf{Total} & \textbf{Running}  & \textbf{Running}  \\
    \ & & & \textbf{Cycles} & \textbf{Total (FP)} & \textbf{Total (INT)} \\ 
      \hline\hline
     \multirow{6}{*}{1} & {Add/Sub} & {16} & {512} & {512} & {-}   \\ 
     \ & {Move} & {2}  & {64}  & {-}  & {64}    \\
    \ & {Complex} & {6}  & {192}  & {704}  & {-}    \\
    \ & {Move} & {1}  & {32}  & {-}  & {96}    \\
    \ & {Sub(FP)} & {1}  & {32}  & {736}  & {-}    \\
    \ & {Cplx} & {4}  & {128}  & {864}  & {-}    \\
\hline
     \multirow{3}{*}{2} & {Add/Sub} & {16} & {512} & {1376} & {-}   \\ 
     \ & {CPLX INT} & {6}  & {192}  & {-}  & {288}    \\
    \ & {Complex} & {2}  & {64}  & {1440}  & {-}    \\
    \hline
     \multirow{2}{*}{3} & {Add/Sub} & {16} & {512} & {1952} & {-}   \\ 
     \ & {Move} & {6}  & {256}  & {-}  & {544}    \\
    \hline
    \multirow{2}{*}{Complex} & {Add (INT)} & {1x7} & {32x7} & {-} & {768}   \\ 
     \ & {Complex} & {6x7}  & {192x7}  & {3296}  & {-}    \\
    \hline
    \ {Total} & {} & {} & {} & {3296} & {768} \\
    \hline
    \end{tabular}
  \end{center}
\end{table}

\begin{table*}
\footnotesize
  \begin{center}
    \caption{eGPU vs. FFT IP Core}
    \label{tab:compare_ip}
    \begin{tabular}{|c||c|c|c|c||c|c|c|c||c|c||}
\hline
  \textbf{FFT Size} & \textbf{IP} & \textbf{ALM} & \textbf{M20K} & \textbf{DSP} & \textbf{eGPU} & \textbf{ALM} & \textbf{M20K}& \textbf{DSP} & \textbf{Ratio}  & \textbf{Ratio} \\
  \ & & \textbf{Registers} & & & & \textbf{Registers} & & & \textbf(Performance) & \textbf{(Normalized)} \\
      \hline\hline
     \ {256} & {0.50$\mu$s} & {12842/23284} & {62} & {32} & {2.54$\mu$s} & {8801/15109} & {192} & {32} & {5.1} & {2.6}  \\
     \hline
     \ {1024} & {1.84$\mu$s} & {15350/25859} & {93} & {40} & {12.65$\mu$s} & {8801/15109} & {192} & {32} & {6.9} & {3.5}   \\
     \hline
     \ {4096} & {6.92$\mu$s} & {18227/31283} & {126} & {48} &  {46.05$\mu$s} & {8801/15109} & {192} & {32}  & {6.7} & {3.3}   \\
    \hline
    \end{tabular}
  \end{center}
\end{table*}

\begin{table}
\footnotesize
  \begin{center}
    \caption{FFT Efficiency - A100 vs. eGPU}
    \label{tab:compare_gpu}
    \begin{tabular}{|c||c|c|c||}
\hline
  \textbf{GPU} & \textbf{256} & \textbf{1024} & \textbf{4096}  \\
  \ & \textbf{points} & {points} & {points} \\
      \hline\hline
     \ {eGPU} & {25\%} & {27\%} & {36\%}  \\
     \hline
     \ {V100} & {15\%} & {18\%} & {21\%}   \\
    \hline
     \ {A100} & {21\%} & {27\%} & {33\%}   \\
    \hline
    \end{tabular}
  \end{center}
\end{table}

\subsection{Radix-8 DFT Analysis}

This function is comprised of three radix-2 passes (all resident in a thread throughout the computation), followed by seven complex rotations. For any pass of the FFT, there is only one butterfly performed, but it will be run on every active thread. We use the example of the 4096 point FFT here; this will require 512 threads (4096/8), or a 32-deep wavefront. We count the number of operations per thread, and the total number of machine cycles required. The total number of operations is 16$\times$ this, as we have 16 SPs running in parallel. 

Each pass starts with 4 complex subtractions and 4 complex additions (16 real FP operations). The additions are stored in place, and the subtractions in temporary registers for rotation. The first subtraction is not rotated, and just stored back in place (2 INT operations). The next value has a complex coefficient applied (6 FP operations for the complex multiply). The third value (0-j), can be calulated using two INT, or one INT and one FP operation. The last value (0.707-0.707j) only requires 4 (rather than 6) FP operations. The subsequent passes use simpler coefficients, which in turn need fewer operators. Finally, seven complex multiplications apply the twiddle factors to each DFT (42 total FP operations). At the end of the pass we have 3296 FP operations, plus 768 INT operations (mostly used for MOVEs). If we include the 288 INT operators which perform simple FP operations, the efficiency of this FFT in the eGPU-DP variant increases to 20.5\%, up from 19.13\%. 

\subsection{Mixed Radix FFT}

All of the points and radix combinations in Tables~\ref{tab:r16_fft},~\ref{tab:r8_fft}, and~\ref{tab:r4_fft} use a single radix, except for the radix-16 1024 point FFT, where the last pass is in radix 4. As the number of threads required for the radix-4 pass are four times as much for the radix 16 passes, we implemented the radix-4 processing in four separate blocks, each processing one quarter of the pass, while reusing the thread initialization for the base radix-16 case.

\section{Comparison Against other Solutions}

\begin{figure*}
    \centering
    \Description{Floorplan}
    \includegraphics[scale=0.40]{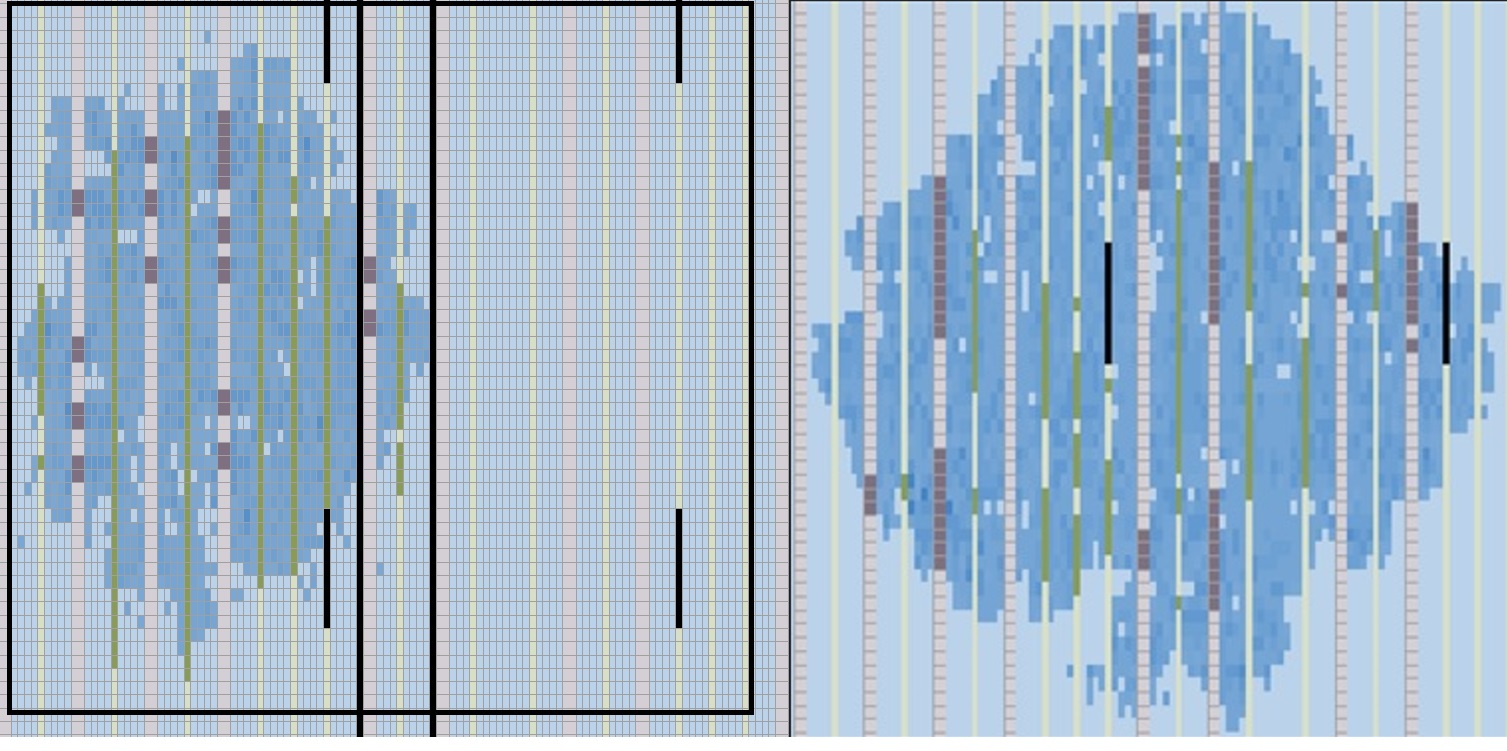}
    \caption{eGPU vs. 4K Streaming FP FFT IP}
    \label{fig:floorplan_compare}
\end{figure*}

We first compare against the Intel Streaming FP32 FFT IP cores~\cite{IntelFFT}. The IP cores are faster, but they are also larger. We tabulate the individual features (ALMs, DSPs, M20Ks) for reference, although we base the cost on the footprint of the respective solution in the FPGA (the ALM cost roughly correlates with the footprint ratio). Figure~\ref{fig:floorplan_compare} compares the two floorplans - eGPU (64KB shared memory) on the left, and the 4K FFT IP core on the right. For clarity, a black box is drawn around the eGPU to show the area required by the IP core. Although the number of DSP Blocks are 50\% greater in the IP core (and the memory blocks are actually less), these resources would be largely unreachable by other parts of the design, as the FFT IP ALM logic is wrapped around them. We can therefore conclude that the FFT IP core is twice the cost of the eGPU instance. We summarize the results in Table~\ref{tab:compare_ip}. The IP core has an absolute performance advantage over the soft GPGPU (using radix-16 FFTs) of almost 7$\times$, but closer to 3$\times$ once the results are normalized for resource cost.

We compare the soft GPGPU to the Nvidia A100 based on efficiency in Table~\ref{tab:compare_gpu} (it would be unfair to compare the soft GPGPU, which only utitlizes $\approx$ 1\% of a mid range FPGA to a large GPGPU on absolute performance). Put another way, the eGPU will occupy an area somewhat less than 1mm$^2$, and the GPU an area of over 800mm$^2$. On the FPGA we can also instantiate many GPGPU cores if needed, and if we wanted absolute performance purely based on FFTs, we could also instantiate multiple FFT cores (or indeed use a much higher performance variant of the FFT cores that supports a much higher bandwidth parallel interface). The Nvidia GPGPU efficiencies (we also show the older V100~\cite{TeslaV100} for interest), are from the most recent cuFFT libraries reported by Nvidia~\cite{cuFFT}.

\section{Conclusions}

We have described and reported on some architectural enhancements to soft GPGPU cores that used the resource (DSP Blocks and embedded memory) types and features that are particular to FPGAs. Building an ASIC-style shared memory bank system in FPGA is very costly because of the logic and routing complexity required, so we created a an alternate higher-bandwidth manually controlled banking approach. We described a complex multiplier enhancement that mitigated wire load, and did not increase the footprint of the soft GPGPU. These combination of these new approaches improved the efficiency of FP FFTs by up to 50\%. As part of our analysis, we reported the profile of 48 different combinations of FFT decomposition, points, and processor architectures, which we hope will serve as a starting point for other investigators.

We set out to see whether a soft processor could be competitive with soft IP on FPGAs. Our results show that there is only about a 3$\times$ advantage of the fixed function soft IP. As both the soft GPU and the IP core occupy in the range of 1\%-2\% of the FPGA device, we can use one or both, or multiple copies of each for our designs. 

The flexibility of the soft GPGPU allows us to implement algorithms that are too difficult or time consuming to create, debug, and modify in RTL. We compared the efficiencies (sustained to peak use of available FP resources) with recent Nvidia GPGPUs running the latest version of the cuFFT programs, and found that our soft GPGPU was able to match or beat their efficiency (measured in the utilization of the FP operators). This shows that a soft GPGPU can be constructed that has a valid architecture, and useful results for the dataset sizes that will likely be used in a FPGA system design.

Finally, we described another way to compare different FPGA designs. Rather than cataloging the resources individually, we look at the overall impact of the design fit, by comparing the floorplan footprint. Evaluating the physical cost of the design will give us a more correct idea of the cost, by considering both used and unreachable resources.

\newpage\clearpage


\end{document}